\documentclass[twocolumn,showpacs,amsmath,amssymb,nofootinbib]{revtex4-1}
\usepackage{dcolumn}
\usepackage{bm}
\usepackage{graphicx}

\newcommand{\be}{\begin{equation}}
\newcommand{\ee}{\end{equation}}
\newcommand{\ba}{\begin{eqnarray}}
\newcommand{\ea}{\end{eqnarray}}

\newcommand{\n}[1]{\label{#1}}
\newcommand{\eq}[1]{Eq.(\ref{#1})}

\newcommand{\hh}{\, ,\hspace{0.5cm}}
\newcommand{\hhh}{\, ,\hspace{0.2cm}}

\newcommand{\BM}[1]{{\mbox{\boldmath $#1$}}}

\newcommand{\ve}{\varepsilon}
\newcommand{\G}{\mathcal{T}}
\newcommand{\dx}{\dot{x}}

\newcommand{\CQG}[4]{{#1} {Classical Quantum Gravity\ }{\bf #2},\  #3 (#4).}

\newcommand{\PRD}[4]{{#1}{Phys. Rev.\  D\ }{\bf #2},\  #3 (#4).}

\newcommand{\PR}[4]{{#1}{Phys. Rev.\ }{\bf #2},\  #3 (#4).}

\newcommand{\AdP}[4]{{#1}{Annalen der Physik\ }{\bf #2},\ #3 (#4).}

\newcommand{\APJ}[4]{{#1}{Astrophys. J.\ }{\bf #2},\  #3 (#4).}

\newcommand{\JETP}[4]{{#1}{JETP\ }{\bf #2},\  #3 (#4).}

\newcommand{\BOOKIN}[5]{{#1,\ }{in {\it #2}}\ {edited by #3\ }{(#4)\ }{p. #5.}}
\newcommand{\BOOK}[4]{{#1}{ {\it #2,\ }}{#3\ }{(#4)}}

\begin{document}

\title{Generalized Fermat's principle and action for light rays in a curved spacetime}
\author{Valeri P. Frolov}
\email{vfrolov@ualberta.ca}
\affiliation{Theoretical Physics Institute, University of Alberta,
Edmonton, AB, Canada,  T6G 2G7}
\date{\today}

\begin{abstract}
We start with formulation of the generalized Fermat's principle for light propagation in a curved spacetime. We apply Pontryagin's minimum principle of the optimal control theory and obtain an effective Hamiltonian for null geodesics in a curved spacetime. We explicitly demonstrate that dynamical equations for this Hamiltonian correctly reproduce null geodesic equations. Other forms of the action for light rays in a curved spacetime are also discussed.
\end{abstract}

\pacs{41.20.Jb, 42.15.-i, 42.81.Gs, 04.20.Cv, 04.70.Bw \hfill}

\maketitle

\section{Introduction}

It is well known  that the Maxwell equations in a curved  spacetime can be written in the form of the electromagnetic field equations in a flat space-time in the presence of a ''medium'' with the dielectric permittivity and magnetic permeability tensors related to the spacetime metric \cite{Pleb,VoIzSk,Mash}. In the geometric optics approximation the problem of propagation of high frequency electromagnetic waves is reduced to study the geometry of null geodesics, that play the role of bi-characteristics for the wave fronts. Thus this problem is similar to ordinary optics in the media. Fermat's principle proved to be a useful tool for study this problem \cite{BoWo}. The Fermat's principle for light propagation in a static spacetime was formulated in \cite{Weyl,Pauli,MTW}. Its generalization to a stationary spacetime was given in the book by Landau and Lifshitz\footnote{The first edition of this book appeared in Russian in 1939} \cite{LL} and later in a nice paper by Brill \cite{Brill}. Fermat's principle basically states, that a light ray from a spatial point 1 to a point 2 propagates along such a path that takes the least time. Recently Kovner \cite{Kovner} proposed a generalization of the Fermat's principle which is valid for light rays in an arbitrary (not necessary stationary) spacetime. According to this principle the arrival time of
a future directed null ray emitted from some point 1 and registered by an observer, moving along any timelike worldline, is stationary (minimal) for a null geodesic.
A simple derivation of this result was given in \cite{NiSa}. A comprehensive discussion of the generalized Fermat's principle can be found in the papers by Perlick \cite{Pe:90a,Pe:90b}. He  also formulated an action for light rays in an arbitrary metric based on the Fermat's principle \cite{Pe:90a}.

In the present paper we propose a slightly different approach for study light rays in a curved  $(n+1)-$dimensional spacetime based on the Fermat's principle. One can foliate the spacetime region by a $n-$parameter family of timelike curves. The coordinates $x^i$, `enumerating' this curves can be used as coordinates in a reduced $n-$dimensional space $\Gamma^{n}$. Any $n-$dimensional curve connecting an initial point $x^i_1$ with an end point $x^i_2$ in $\Gamma^{n}$ can be uniquely lifted as a future directed null curve in $(n+1)-$dimensional spacetime $M^{n+1}$, provided one fix time $x^0_1$ of its emission.
Time $x^0_2$ of the null ray arrival to the end point $x_2^i$ is a functional of the spatial path. Finding minimal value of this functional is a standard problem of the optimal control theory (see e.g. \cite{PePl,SuWi,Young}. To solve this problem one can use the Pontryagin's minimal principle \cite{Pont}. In this paper we use this approach to construct the effective Hamiltonian, and demonstrate that corresponding dynamical equations are equivalent to the equations derived from Perlick's action \cite{Pe:90a} and correctly reproduce null geodesic equations in the spacetime $M^{n+1}$.

\section{Motivation}

An action for a particle motion in an external gravitational field in the spacetime $M^{n+1}$ can be written in the form
\ba\n{part_action}
S&=&\int_{\lambda_1}^{\lambda_2} d\lambda L\, ,\\
L&=&{1\over 2}\left[ \eta^{-1}g_{\mu\nu}{dx^{\mu}\over d\lambda}
{dx^{\nu}\over d\lambda}-m^2 \eta\right]\, .
\ea
Here $\mu,\nu=0,1,\ldots, n$, $g_{\mu\nu}$ is the metric, $x^{\mu}(\lambda)$ is a path of a particle and $m$ is its mass. $\eta(\lambda)$ is a Lagrange multiplier, which under the change of the parametrization $\lambda\to \tilde{\lambda}(\lambda)$ transforms as follows
\be
\tilde{\eta} d\tilde{\lambda}=\eta d\lambda\, ,
\ee
so that the action \eq{part_action} is reparametrization invariant. The variation of the action is
\ba
\delta S&=&-\int_{\lambda_1}^{\lambda_2} d\lambda \left[
{ D\over d\lambda }\left( \eta^{-1} {dx^{\mu}\over d\lambda}\right)\delta x_{\mu}\right.\nonumber\\
&+&\left. {1\over 2}\left(\eta^{-2}g_{\mu\nu}{dx^{\mu}\over d\lambda}
{dx^{\nu}\over d\lambda}+m^2\right)\delta\eta\right]\\
&+& \left. \left( \eta^{-1}g_{\mu\nu}{dx^{\nu}\over d\lambda} \delta x^{\mu}\right) \right|_{\lambda_1}^{\lambda_2}\, .
\ea
Here $D/ d\lambda$ is a covariant derivative in the direction $dx^{\mu}/ d\lambda$.
For an extremum  of the action the following conditions are to be satisfied:
\ba
&&{ D\over d\lambda }\left( \eta^{-1} {dx^{\mu}\over d\lambda}\right)=0\, ,\n{geq}\\
&&g_{\mu\nu}{dx^{\mu}\over d\lambda}
{dx^{\nu}\over d\lambda}=-\eta^{2}m^2\, ,\n{cond}\\
&&\left. \left( \eta^{-1}g_{\mu\nu}{dx^{\mu}\over d\lambda} \delta x_{\mu}\right) \right|_{\lambda_1}^{\lambda_2}=0\, .\n{nat}
\ea
$\eta(\lambda)$ is an arbitrary function, depending on the choice of the parametrization. After variations, one can always put it equal to 1. For this choice one gets an affine parametrization. The last condition determines, what is called the `natural boundary conditions' \cite{Cour,Gelf}.

Let us compare cases of massive and massless particles. To illustrate an important difference between these two cases let us assume that the spacetime is 4-dimensional and flat, that is $g_{\mu\nu}=\eta_{\mu\nu}=\mbox{diag}(-1,+1,1,1)$. Let us fix an initial point $x^{\mu}(\lambda_1)=x_1^{\mu}$. For the massive case the conditions (\ref{geq}) and (\ref{cond}) imply that a curve is timelike geodesic, that is, it is a  straight line in the interior of the future null cone $N$ with the vertex at $x_1^{\mu}$.
It is evident that for any choice of the end point $x^{\mu}(\lambda_2)=x_2^{\mu}$ within $N$ there exist only one timelike geodesic connecting it with $x_1^{\mu}$. This means that for $m\ne 0$ a variational problem with fixed end points is well defined and has a unique solution.

\begin{figure}
  \includegraphics[width=7cm]{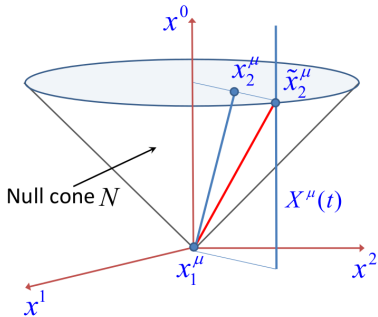}\\
  \caption{Illustration to variation problem for
  geodesics in a flat spacetime. $x_1^{\mu}$ is a fixed initial point.
  For a massive particle for any end point $x_2^{\mu}$ inside the null cone $N$ there exists a unique solution. For a massless case the end point must lie on the surface of the null cone $N$ (point $\tilde{x}_2^{\mu}$). For such a problem one can only require that the end point of the geodesic is at some timelike curve $X^{\mu}(t)$.}\label{Fig}
\end{figure}

In the massless case the situation is quite different.
The conditions (\ref{geq}) and (\ref{cond}) imply that a curve is a null geodesic, and hence it is a generator of the null cone $N$ emitted from $x_1^{\mu}$. A solution of the problem with a fixed end point exists only when the end point is on this null cone $N$, that is on a special submanifold of dimension 3. The measure of this submanifold is zero. Hence, for a generic choice of the end point $x^{\mu}(\lambda_2)=x_2^{\mu}$ a solution of the variational problem with fixed end points does not exist. In order to have a well defined variational problem for the massless case one may allow the end point not to be fixed but to belong, for example, to a timelike curve. The simplest choice is a straight line
\be
X^{\mu}(\tau)=V^{\mu} \tau +B^{\mu} +x^{\mu}_1\hh \BM{V}^2=-1\, .
\ee
This is a trajectory of an observer moving with constant 4-velocity $V^{\mu}$. It crosses the future null cone $N$ emitted at $x^{\mu}_1$ at his/her proper time $\tau_+$
\be
\tau_+=\sqrt{\BM{B}^2+(\BM{V},\BM{B})^2}+(\BM{V},\BM{B})\, .
\ee
It is evident that a similar modification of the boundary value problem for the massive particle, when the end point belongs to a timelike line, does not give a unique solution. Any point of $X^{\mu}(\tau)$ for $\tau>\tau_+$ can be connected by a timelike geodesic with the initial point $x_1^{\mu}$. Hence, instead of a unique solution one would have a one-parameter family of such solutions. To summarize, the variational problems for massive and massless particles are quite different, and one cannot obtain the latter by simply taking the limit $m\to 0$ in the action.

\section{Generalized Fermat's principle}

Let us consider variational problem for null rays in an arbitrary $(n+1)-$dimensional curved spacetime $M^{n+1}$. We shall not deal with cases when there is more than one geodesics connecting two chosen point. This and other global aspects of the problem are beyond the scope of this paper. In this sense, the following consideration is local.
Consider  a $n-$dimensional set of timelike curves $x^{\mu}=x^{\mu}(\tau,x^i)$, where a line is specified by a condition $x^i=\mbox{fixed}$ and $\tau$ is a proper time parameter along the curve. We assume that only one line of this set passes through each point in the domain under the consideration. In other words one has a foliation of this domain by $n-$dimensional family of timelike curves.
We denote by $V^{\mu}$  vectors tangent to these curves $V^{\mu}=\partial x^{\mu}/\partial\tau$. In the coordinates $(\tau,x^i)$ the spacetime metric is of the form\footnote{If instead of the proper time $\tau$ one uses any time coordinate  $x^0$ a metric takes the form
\ba
ds^2&=&h dS^2\hh dS^2=-(dx^0-g_i dx^i)^2+\gamma_{ij}dx^i dx^j\, ,\nonumber\\
h&=&-g_{00}\hh g_i=g_{oi}/h\hh \gamma_{ij}=g_{ij}/h+ g_{0i}g_{0j}/h^2\, .\nonumber
\ea
Since equations for null lines and null geodesics are conformally invariant, one can use the conformal metric $dS^2$ for study these equations, so that even for an arbitrary choice of the coordinate time $x^0$ the problem, after this transformation, effectively reduces to a similar problem in the metric \eq{genmetric}}
\be\n{genmetric}
dS^2=-(d\tau-g_i dx^i)^2+\gamma_{ij}dx^i dx^j\, .
\ee
We assume that Greek indices, $\mu, \nu, \ldots$ take values $0,1,\ldots n$, while the Latin indices, $i,j,\ldots$ take values $1,2,\ldots n$.
Consider a null curve $x^{\mu}(\lambda)$ and denote by $l^{\mu}$ a tangent vector to it,
$l^{\mu}=dx^{\mu}/d\lambda=(\dot{x}^0,\dot{x}^i)$. We denote by a dot a derivative of a function of $\lambda$ with respect to its argument. A condition that the curve is null, $l_{\mu}l^{\mu}=0$, takes the form
\be
-(\dot{\tau}-g_i \dot{x}^i)^2+\gamma_{ij}\dot{x}^i \dot{x}^j=0\, .
\ee
A solution of this equation for $\dot{\tau}>0$ is
\be
\dot{\tau}=g_i \dot{x}^i + U\hh
U=\sqrt{\gamma_{ij}\dot{x}^i \dot{x}^j}\, .\n{lift}
\ee
Let $x^i=x^i(\lambda)$ be a curve in space $\Gamma^n$ with coordinates $x^i$.
Equation (\ref{lift}) allows one to lift it to the spacetime $M^{n+1}$ as a null curve $x^{\mu}(\lambda)$. It is sufficient to choose the value of $\tau$ at the initial point, $\tau(\lambda_1)=\tau_1$, and solve ODE (\ref{lift}) with this initial data.

Consider a 1-parameter family of  curves $x^i(\lambda,\ve)$ connecting an initial point $x^i_1$ and a final point $x^i_2$ in the space $\Gamma^{n}$, so that
\be
x^i(\lambda_1,\ve)=x^i_1\hh
x^i(\lambda_2,\ve)=x^i_2\, .
\ee
We consider the corresponding 1-parameter family of lifted curves $x^{\mu}(\lambda,\ve)$ in the spacetime $M^{n+1}$ that have the same initial value $x^0(\lambda_1,\ve)=\tau_1$.
We choose this family so that the lifted $(n+1)-$dimensional curve for $\ve=0$ is a null geodesic.
For a given curve $x^i(\lambda,\ve)$ from the family, the final value $x^0(\lambda_2,\ve)=\tau_2(\ve)$ can be found by solving the equation \eq{lift}.

Let us show now that the arrival time $\tau_2(0)$ for a null geodesic gives an extremum for a function $\tau_2(\ve)$. Let us fix the initial point $x^{\mu}=(\tau_1,x^i_1)$ and the final $n-$dimensional point $x^i_2$. Then the action
\eq{part_action} is a function of $\ve$, $S(\ve)$. When equations (\ref{geq}) and (\ref{cond}) are satisfied, the variation of this function takes the form
\be
\delta_{\ve} S=\left. \eta^{-1}g_{\mu\nu}\dx^{\nu}\delta x^{\mu}\right|_2=
\eta_2^{-1}(l_i \delta x_2^i+l_{\tau} \delta \tau_2)\, .
\ee
Since $\delta x_2^i=0$ and $\eta^{-1}l_{\tau}\ne 0$, the natural boundary conditions (\ref{nat}) imply that
\be
\left. {\partial \tau_2(\ve)\over \partial \ve}\right|_{\ve=0}=0\, .
\ee
To summarize, the light emitted at $x^{\mu}_1$ reaches an observer at $x^i_2$ earlier than any other null curve. This is an evident generalization of the Fermat's principle for a general case of a curved spacetime.

\section{Optimal control approach, Pontryagin's minimum principle, and an effective Hamiltonian}

Using the generalized Fermat's principle one can formulate the problem of finding light rays in the following way. Consider the following set of first order equations
\ba\n{dyneq}
{dx^0\over d\lambda}&=& \mathcal{Q}(x^{\mu},u^i)\, ,\n{dy1}\\
{dx^i\over d\lambda}&=&u^i\, \n{dy2}\\
\mathcal{Q}(x^{\mu},u^i)&=&g_i(x^{\mu})u^i+\sqrt{\gamma_{ij}(x^{\mu}) u^i u^j}\, .\n{dy3}
\ea
We denote by $J$ the following functional
\be\n{JJ}
J=\int_{x_1^{\mu}}^{x_2^i} d\lambda  \mathcal{Q}(x^{\mu}(\lambda),u^i(\lambda)\, .
\ee
Here and later we use the notation $\int_{x_1^{\mu}}^{x_2^i} d\lambda\ldots$ to indicate that we fix at $\lambda_1$ an initial spacetime point $x_1^{\mu}$, while at the end point $\lambda_2$ we fix only its spatial position $x_2^i$.

The value of $J$ calculated for a solution of (\ref{dy1})-(\ref{dy3}) coincides with $\delta\tau=\tau_2-\tau_1$. According to the generalized Fermat's principle the light ray trajectory provides an extremum (minimum) of the functional $J$.

One may consider such a task as a version of a general optimal control problem, by identifying $u^i(\lambda)$ with control functions. To solve this problem one introduces the following Hamiltonian \cite{Pont,Hull}
\be\n{eh}
\mathcal{H}(\psi_{\mu},x^{\mu},u^i)=\psi_0 \mathcal{Q}+\psi_i u^i\, .
\ee
The corresponding Hamilton equations are
\ba
{d\psi_{\mu}\over d\lambda}&=&-{\partial\mathcal{H}\over \partial x^{\mu}}=-\psi_0 {\partial{\mathcal{Q}}\over \partial x^{\mu}}\, ,\n{pp1}\\
{dx^0\over d\lambda}&=&{\partial\mathcal{H}\over \partial \psi_0}=\mathcal{Q}\, ,\n{pp2}\\
{dx^i\over d\lambda}&=&{\partial\mathcal{H}\over \partial \psi_i}=u^i\, .\n{pp3}
\ea
According to Pontryagin's minimum principle one should substitute a solution of these equations for given control functions $u^i$ into the effective Hamiltonian $\mathcal{H}$, and after this determine them from the condition that they minimize the value of the Hamiltonian. In our case, when $u^i$ are smooth functions, the corresponding conditions are
\be\n{pp4}
{\partial\mathcal{H}\over \partial u^i}\equiv \psi_0 {\partial \mathcal{Q}\over \partial u^i} +\psi_i=0\, .
\ee
When satisfying along the trajectory these equations are necessary condition for an optimum.

A set of equations (\ref{pp1})--(\ref{pp4}) can be simplified. In particular, \eq{pp4} gives
\be
\psi_i=-\psi_0 {\partial \mathcal{Q}\over \partial u^i}\, ,
\ee
while the equations (\ref{pp1})-(\ref{pp3}) can be written in the form
\ba
&&{d \over d\lambda}\left(\psi_0 {\partial {Q}\over \partial\dot{x}^i}\right)-\psi_0 {\partial {Q}\over \partial {x}^i}=0\, ,\n{pp5}\\
&&{d \psi_0\over d\lambda}=-\psi_0 {\partial {Q}\over \partial{x}^0}\, ,\n{pp6}\\
&&{d x^0\over d\lambda}=Q\, .\n{pp7}
\ea
Here we denote
\ba
Q&=&Q(x^{\mu},\dot{x}^i)=\mathcal{Q}(x^{\mu}, u^i=\dot{x}^i)\\
&=&g_i\dot{x}^i+U\hh U=\sqrt{\gamma_{ij}\dot{x}^i \dot{x}^j}\, .\n{pp8}
\ea
Using \eq{pp6} to exclude $\psi_0$ from \eq{pp5} and taking into account \eq{pp7} one obtains the following relation
\be\n{QQ}
{d \over d\lambda}\left({\partial {Q}\over \partial\dot{x}^i}\right)- {\partial {Q}\over \partial{x}^i}
-{\partial {Q}\over \partial {x}^0}{\partial {Q}\over \partial\dot{x}^i}=0\, .
\ee
Equations (\ref{QQ}) and (\ref{pp7}) totally determine a null ray trajectory.
It is easy to check that these equations are reparametrization invariant, that is they are invariant under the transformation $\lambda\to \tilde{\lambda}=f(\lambda)$. We shall now explicitly demonstrate that they are equivalent to the $(n+1)-$dimensional geodesic equations for null rays.

\section{Consistency}

\subsection{Null geodesics in $M^{n+1}$}

Consider a null ray $x^{\mu}(\lambda)$ and denote, as earlier, $l^{\mu}=\dot{x}^{\mu}$ a tangent vector to it. In the metric (\ref{genmetric}) one has
\ba
l^{\mu}&=&(U+g_i\dot{x}^i,\dot{x}^i)\, ,\\
l_{\mu}&=&(-U, g_i U+\gamma_{ij}\dot{x}^j)\, .
\ea
The null geodesic equation in $M^{n+1}$, (\ref{geq}), has the form
\be\n{Dll}
{Dl^{\mu}\over d\lambda}\equiv l^{\nu} l^{\mu}_{;\nu}=\tilde{F} l^{\mu}
\hh \tilde{F}=\eta^{-2}\dot{\eta}
\, .
\ee
Here, as earlier, $D/d\lambda$ is the covariant derivative in $l^{\mu}$ direction. We do not assume that $\lambda$ is an affine parameter, so that $\tilde{F}$ does not necessary vanishes.

Let us consider first the relation
\be\n{ng0}
g_{0\mu}{Dl^{\mu}\over d\lambda}\equiv {Dl_{0}\over d\lambda}=\tilde{F} l_0\, .
\ee
This equation gives
\be\n{UUU}
{dU\over d\lambda}+\Gamma^{\alpha}_{0\beta}l_{\alpha}l^{\beta}=\tilde{F} U\, .
\ee
The expressions for the Christoffel symbols for the metric \eq{genmetric} are given in the Appendix. After simplifications one obtains the following relation
\be\n{UU}
{dU\over d\lambda}+U g_{i,0}\dot{x}^i+{1\over 2} \gamma_{ij,0}\dot{x}^i\dot{x}^j-\tilde{F}U=0\, .
\ee

The other null geodesic equations can be written in the form
\be\n{ng3}
\gamma_{ij}{Dl^{j}\over d\lambda}=\tilde{F}_i
\hh \tilde{F}_i=\tilde{F} \gamma_{ij}\dot{x}^j
\, .
\ee
They can be rewritten in the following explicit form
\ba
&&\gamma_{ij}{\mathcal{D}^2x^j\over d\lambda^2}+\tilde{B}^{(0)}_i+\tilde{B}^{(1)}_i U+\tilde{B}^{(2)}_i U^2=\tilde{F}_i\, ,\n{Deq}\\
&&\tilde{B}^{(0)}_i=g_{i,0}(g_k\dot{x}^k)^2+g_{ij,0}\dot{x}^j g_k \dot{x}^k-{1\over 2} g_i g_{jk,0}\dot{x}^j\dot{x}^k\, ,\nonumber\\
&&\tilde{B}^{(1)}_i=2g_{i,0}g_k \dot{x}^k+g_{ij,0}\dot{x}^j+2A_{ij}\dot{x}^j\, ,\n{NULL}\\
&&\tilde{B}^{(2)}_i=g_{i,0}\, .\nonumber
\ea
We denote by
${\mathcal{D}/ d\lambda}$ a
$n-$dimensional  covariant derivative with respect the metric $\gamma_{ij}$ in the direction $\dot{x}^i$
\ba
&&{\mathcal{D}^2 x^j\over d\lambda^2}={d^2 x^j\over d\lambda^2}+
\G^{i}_{jk} \dx^j \dx^k\, ,\\
&&\G_{i,jk}={1\over 2}(\gamma_{ij,k}+\gamma_{ik,j}-\gamma_{jk,i})\, ,\\
&&\G^{i}_{jk}=\gamma^{in}\G_{n,jk}\, .\nonumber
\ea

\eq{UU} follows from \eq{NULL} and the relation
\be\n{xdot}
\dx^0=g_i \dx^i+U\, .
\ee
To show this it is sufficient to multiply \eq{Deq} by $\dot{x}^i$ and use the relation
\be
\gamma_{ij}\dx^i {\mathcal{D}^2x^j\over d\lambda^2}=U{dU\over d\lambda} -{1\over 2}\gamma_{ij,0}\dot{x}^0\dot{x}^i\dot{x}^j\, .
\ee

Thus, equations (\ref{Deq}), (\ref{NULL}) and (\ref{xdot}) uniquely determine a null geodesic in $M^{n+1}$.

\subsection{Hamiltonian equations}

Let us now write \eq{QQ} in an explicit form. We notice that
\ba
&&{\partial Q\over \partial \dot{x}^i}= g_i+{1\over U} \gamma_{ij}\dot{x}^j\, ,\\
&&{\partial Q\over \partial {x}^i}=g_{j,i}\dx^j+{1\over 2U}\gamma_{kl,i}\dx^k \dx^l\, ,\\
&&{\partial Q\over \partial {x}^0}=g_{i,0}\dx^i +{1\over 2U}\gamma_{kl,0}\dx^k \dx^l\, ,\\
&&{d\over d\lambda}{\partial Q\over \partial \dot{x}^i}=g_{i,0}(g_k\dx^k +U)+
g_{i,j}\dx^j\nonumber\\
&&-{1\over U^2}{dU\over d\lambda}\gamma_{ij}\dx^j+{1\over U}\gamma_{ij,0}\dx^j(g_k\dx^k+U)\\
&&+{1\over U}(\gamma_{ij,k}\dx^j \dx^k+\gamma_{ij}\ddot{x}^j)\, .\nonumber
\ea
Using these relations and \eq{pp7} one can show that \eq{QQ} is equivalent to the following equations
\ba
&&\gamma_{ij}{\mathcal{D}^2x^j\over d\lambda^2}+B^{(0)}_i+B^{(1)}_i U+B^{(2)}_i U^2=F_i\, ,\n{QQQ1}\\
&&B^{(0)}_i=g_k\dx^k\gamma_{ij,0}\dx^j-{1\over 2}g_i\gamma_{jk}\dx^j \dx^k
\, ,\nonumber\\
&&B^{(1)}_i=g_{i,0} g_k\dx^k +\gamma_{ij,0}\dx^j-g_i g_{k,0}\dx^k
+2A_{ij}\dot{x}^j\, ,\n{QQQ}\\
&&B^{(2)}_i=g_{i,0}\hh F_i=F\gamma_{ij}\dx^j\, ,\nonumber\\
&&F={1\over U}{dU\over d\lambda}-g_{k,0}\dx^k -{1\over 2 U}\gamma_{kl,0}\dx^k \dx^l\, .\n{QQQ2}
\ea
By comparing \eq{NULL} and \eq{QQQ} and using relation $g_{ij}=\gamma_{ij}-g_i g_k$ one can see that the coefficients in equations (\ref{Deq}) and (\ref{QQQ1}) are identical, namely
\be
\tilde{B}^{(0)}_i=B^{(0)}_i\hh \tilde{B}^{(1)}_i=B^{(1)}_i\hh \tilde{B}^{(2)}_i=B^{(2)}_i\, .
\ee
To have identical right hand sides of this equations one must put $\tilde{F}=F$. This condition guarantees that in the both cases the parametrization $\lambda$ is identical.

To summarize, we demonstrated that equations (\ref{QQ}) and (\ref{pp7}),   obtained from the Hamiltonian \eq{eh}, are equivalent to the equations (\ref{Dll}) for a null geodesic in $M^{n+1}$.

\section{Perlick's form of the action}

Perlick \cite{Pe:90a} proposed an elegant form of the action for null rays. Its slightly modified version is
\ba
&&\mathcal{S}[x^{\mu}(\lambda),\mu(\lambda)]=\int_{x_1^{\mu}}^{x_2^i} d\lambda\  \mathcal{L}\, ,\n{ss1}\\
&&\mathcal{L}=\mu(\lambda) (\dx^0-Q)\, ,\n{SS}\\
&&Q=g_i \dx^i +U\hh U=\sqrt{\gamma_{ij}\dx^i dx^j}\, .
\ea
Here $\mu(\lambda)$ is a Lagrange multiplier. The action is invariant under the reparametrization $\lambda\to \tilde{\lambda}(\lambda)$, provided $\mu$ is not transformed, $\tilde{\mu}=\mu$.
Variations of this action give the following equations
\ba
{\delta \mathcal{S}\over \delta \mu}&\equiv& \dx^0-Q=0\, ,\n{na1}\\
-{\delta \mathcal{S}\over \delta x^0}&\equiv& \dot\mu+\mu {\partial Q\over \partial x^0}=0\, ,\n{na2}\\
{\delta \mathcal{S}\over \delta x^i}&\equiv& {d\over d\lambda}\left(\mu {\partial Q\over \partial \dx^i}\right) -\mu {\partial Q\over \partial x^i}=0\, .\n{na3}
\ea
It is easy to see that after putting $\mu=\psi_0$ this set of equations  is identical to equations (\ref{pp5})-(\ref{pp7}) obtained from the effective Hamiltonian (\ref{eh}). Hence, as it was shown in the previous section, it correctly reproduces the null geodesic equations in $M^{n+1}$.

Let us denote by $\{\pi, p_{\mu}\}$ momenta conjugated to $\{\mu,x^{\mu}\}$
\ba
\pi&=&{\partial \mathcal{L}\over \partial\dot{\mu}}=0\, ,\n{pi1}\\
p_0&=&{\partial \mathcal{L}\over \partial\dx^0}=\mu\, , \n{pi2}\\
p_i&=&{\partial \mathcal{L}\over \partial\dx^i}=-\mu{\partial Q\over \partial \dx^i}\, .\n{pi3}
\ea
The Hamiltonian
\be
H^*=\pi\dot{\mu}+p_{\mu} dx^{\mu}-\mathcal{L}
\ee
identically vanishes. This is a consequence of the reparametrization invariance of the action (\ref{ss1}).

Denote
\be
\pi_i={p_i\over p_0}\, .
\ee
Then one has
\be
\pi_i=g_i+{1\over U}\gamma_{ij}\dx^j\, .
\ee
It is easy to check that in addition to a trivial constraint (\ref{pi1})
the following constraint is satisfied
\be
\varphi\equiv {1\over 2}\left[ \gamma^{ij}(\pi -g_i)(\pi_j-g_j)-1\right]=0 \, .\n{cc2}
\ee
These two constraints, (\ref{pi1}) and (\ref{cc2}), are of the first order, that is they Poisson commute.
Following the general theory of dynamical systems with constraints \cite{Dirac} one should introduce an effective Hamiltonian, which is a linear combination of the constraints
\be\n{hhh}
H=v \pi +u \varphi\, .
\ee
Here $v=v(\lambda)$ and $u=u(\lambda)$.

One has
\ba
&&{\partial \varphi\over \partial p_0}={\partial \varphi\over \partial \pi_i} {\partial \pi_i\over \partial p_0}=-{1\over p_0}\pi_i {\partial \varphi\over \partial \pi_i}\, \\
&&{\partial \varphi\over \partial p_i}={\partial \varphi\over \partial \pi_i} {\partial \pi_i\over \partial p_i}={1\over p_0} {\partial \varphi\over \partial \pi_i}\, \\
&&{\partial \varphi\over \partial \pi_i}=\gamma^{ij}(\pi_i-g_i)\, ,\\
&&\pi_i {\partial \varphi\over \partial \pi_i}=1-g_i g^i+g^i\pi_i \, .
\ea

The Hamilton equations for \eq{hhh} are
\ba
\dot{\mu}&=& {\partial H\over \partial \pi}=v\, ,\n{HH1}\\
\dot{\pi}&=& -{\partial H\over \partial \mu}=0\, ,\n{HH2}\\
\dx^0 &=& {\partial H\over \partial p_0}=-{u\over p_0} \pi_i {\partial \varphi\over \partial \pi_i}=-{u\over p_0}\pi_i \gamma^{ij}(\pi_i-g_i)\, \n{HH3}\\
\dx^i &=& {\partial H\over \partial p_i}={u\over p_0} {\partial \varphi\over \partial \pi_i}={u\over p_0} \gamma^{ij}(\pi_i-g_i)\, \n{HH4}\\
\dot{p}_0&=& -{\partial H\over \partial x^0}=-u {\partial \varphi\over \partial x^0}\, ,\n{HH5}\\
\dot{p}_i&=& -{\partial H\over \partial x^i}=-u {\partial \varphi\over \partial x^i}\, .\n{HH6}
\ea
\eq{HH2} implies
\be
\pi_i={p_0\over u}\gamma_{ij} \dx^j+g_i\, .
\ee
The constraint equation $\varphi=0$ takes the form
\be
-{u\over p_0}=U\, .
\ee
Thus \eq{HH3} can be written in the form
\be\n{x00}
\dot{x}^0=g_i \dx^i +U\, .
\ee
By combining \eq{HH5} and \eq{HH6} one obtain
\be\n{PP1}
\dot{\pi}_i=U\left({\partial \varphi\over \partial x^i} +\pi_i {\partial \varphi\over \partial x^0} \right)\, .
\ee
Simple calculations give\footnote{It should be emphasized that partial derivatives with respect $x^{\mu}$ in (\ref{HH5}), (\ref{HH6}) and other similar relations are taken for fixed value $p_{\mu}$.}
\be
{\partial \varphi\over \partial x^{\mu}}=-{1\over 2 U^2}\gamma_{jk,\mu}\dx^j \dx^k +{1\over U} g_{j,\mu} \dx^j\, .
\ee
Straightforward calculations allows one to show that the equations (\ref{PP1}) are equivalent to equations (\ref{QQQ1})-(\ref{QQQ2}).

\section{Fermat's principle in a stationary spacetime}

In a stationary spacetime the above considerations are greatly simplified.
Let us write the line element in a stationary spacetime with the Killing vector $\xi^{\mu}$ in the form
\ba
ds^2&=&h dS^2\hh h=-g_{tt}=-\BM{\xi}^2\, ,\n{conf}\\
dS^2&=&-(dt-g_i dx^i)^2+\gamma_{ij} dx^i dx^j\, .\n{ultra}
\ea
Here
\be
g_i={g_{ti}\over h}\hh
\gamma_{ij}={g_{ij}\over h}+g_i g_j\, .
\ee
The expression for $dS^2$ is similar to \eq{genmetric} with only one important difference: its coefficient now do not depend on time $t$. We call the metric \eq{ultra} ultrastationary. The ultrastationary metric $dS^2$ is connected with the physical metric $ds^2$ by the conformal transformation. Since both null lines and null geodesics are conformal invariant, one can always reduce the problem in a stationary spacetime to the corresponding  problem in the ultrastationary one.

The equation \eq{QQ} in the metric \eq{ultra} takes the form
\be
{d \over d\lambda}\left({\partial {Q}\over \partial\dot{x}^i}\right)- {\partial {Q}\over \partial{x}^i}=0\, .
\ee
This is evidently the Euler-Lagrange equation for the action
\be\n{statfer}
S=\int_{x_1^{\mu}}^{x^{i}_2} d\lambda (g_i\dx^i +U)\, . \ee
If null ray is emitted in $x^i_1$ at $t_1$, and arrives to $x^i_2$ at $t_2$,
then this action is nothing but the difference  $t_2-t_1$. The condition that that the action is minimal is nothing but a standard Fermat's principle. This result for stationary spacetimes is well known. Its discussion can be found, for example in \cite{LL,Brill}.

\acknowledgments

The author is grateful to the Natural Sciences and Engineering
Research Council of Canada and the Killam Trust for their financial support.
A part of this work was done during the
Peyresq Physics 18 meeting (June 2013). The author is grateful to its participants for the discussions, and  OLAM and Association pour la Recherche Fondamentale (Bruxelles) for their support.

\appendix

\section{Useful formulas}

For the line element
\be
dS^2=-(d\tau-g_i dx^i)^2+\gamma_{ij}dx^i dx^j\, .
\ee
the metric $g_{\mu\nu}$ and its inverse  $g^{\mu\nu}$ are
\ba
g_{00}&=&-1\hh g_{0i}=g_i\hh g_{ij}=\gamma_{ij}-g_i g_j\, ,\n{metr}\\
g^{00}&=&-(1-g_i g^i)\hh g^{0i}=g^i\hh g^{ij}=\gamma^{ij}\, .\n{invmetr}
\ea
We denote by $\gamma^{ij}$ the inverse of the $n-$dimensional  metric $\gamma_{ij}$.

We define the  Christoffel symbols in $M^{n+1}$ and $\Gamma^n$ as follows
\ba
\Gamma_{\lambda\mu\nu}&=&{1\over 2}(g_{\mu\lambda,\nu}
+g_{\nu\lambda,\mu}-g_{\mu\nu,\lambda})\hhh
\Gamma^{\lambda}_{\mu\nu}=g^{\lambda\alpha}\Gamma_{\alpha\mu\nu}\, ,\\
\mathcal{T}_{kij}&=&{1\over 2}(\gamma_{ik,j}+\gamma_{jk,i}-\gamma_{ij,k})\hhh
\mathcal{T}^k_{ij}=\gamma^{kl}\mathcal{T}_{kij}\, .
\ea
The components of the  Christoffel symbols in $M^{n+1}$ are
\ba
\Gamma_{000}&=&0\hh \Gamma_{00i}=0\hh\Gamma_{i00}=g_{i,0}\, ,\\
\Gamma_{i0j}&=&{1\over 2} g_{ij,0}+A_{ij}\hh
\Gamma_{0ij}=-{1\over 2} g_{ij,0}+S_{ij}\, ,\\
\Gamma{kij}&=&\mathcal{T}_{kij}+g_i A_{jk}+g_j A_{ik}-g_k S_{ij}\, .
\ea
\ba
\Gamma^0_{00}&=&g^i g_{i,0}\hh \Gamma^i_{00}=\gamma^{ij} g_{j,0}\, ,\\
\Gamma^0_{0i}&=&{1\over 2}g^j g_{ij,0}+g^j A_{ji}\, ,\\
\Gamma^j_{0i}&=&\gamma^{jk}({1\over 2} g_{ik,0}+A_{ki})\, ,\\
\Gamma^0_{ij}&=&(1-g_k g^k) ({1\over 2} g_{ij,0}+S_{ij})+ g^k \mathcal{T}_{kij}\nonumber\\
&+&g^i g^k A_{jk} +g_j g^k A_{ik}-g_k g^k S_{ij}\, ,\\
\Gamma^{k}_{ij}&=&\mathcal{T}^k_{ij}-{1\over 2}g^k g_{ij,0}+g_i A_j^{\ k}+g_j A_i^{\ k}\, .
\ea
We denoted
\be
A_{ij}={1\over 2}(g_{i,j}-g_{j,i})\hh
S_{ij}={1\over 2}(g_{i,j}+g_{j,i})\, .
\ee

\end{document}